\documentclass[12pt]{article}
\title{\large Observation of a common symmetry for the pseudogap and the superconducting
order parameter near the surface of underdoped $YBa_2Cu_3O_{6+x}$}

\author{G. Koren, L. Shkedy and E. Polturak\\
 Physics Department, Technion - Israel Institute of Technology\\
 Haifa, 32000, ISRAEL}

\date{ }
\def\bfig {\begin{figure}[tbhp] \centering}
\def\efig {\end{figure}}

\usepackage{amssymb}
\usepackage[pdftex]{graphicx}

\begin{document}
\DeclareGraphicsExtensions{.jpg,.mps,.eps,.pdf}

\normalsize \baselineskip=5mm \sf \maketitle \vspace{25mm}

PACS: 74.20.Rp, 74.50.+r, 74.72.Bk\\ \\

Measurements of the angular dependence of conductance spectra in
the {\em a-b} plane of underdoped $YBa_2Cu_3O_{6+x}$ junctions are
reported. At zero magnetic field the superconducting gap shows a
$|d+is|$-like symmetry. Application of a magnetic field strongly
suppresses this gap leaving only the pseudogap feature which also
shows a $|d+is|$-like angular dependence. We thus observe the same
symmetry for the superconducting gap and the pseudogap
characterizing the YBCO electrodes near the interface with the
barrier. An $H_{c2}$ value of $\sim$5T of the secondary ($is$)
order parameter can also be deduced from our results.

\pagebreak

The concept of the pseudogap in the context of high temperature
superconductivity was first suggested by Alloul {\em et al.} who
measured nuclear magnetic resonance (NMR) of $Y$ in
$YBa_2Cu_3O_{6+x}$ (YBCO).\cite{Alloul} Later, the presence of the
pseudogap was demonstrated by many different techniques, for
example in measurements of angular resolved photo-emission
(ARPES),\cite{Ding} transport studies,\cite{Cooper} tunneling
measurements,\cite{renner} and conductance spectroscopy in
junctions.\cite{DeutscherNature} A comprehensive review of
experimental studies of the pseudogap in the HTS materials was
given by Timusk and Statt.\cite{Timusk} In particular, Deutscher
had shown the existence of two distinct energy scales (energy
gaps) in the high temperature superconductors
(HTS).\cite{DeutscherNature} One is the ordinary superconducting
gap $\Delta$ which exists at $T\le T_c$, while the other is the
pseudogap which opens up at $T^*$ or $T_p$,  both above $T_c$.
Recently, a pseudogap signature was observed also in the electron
doped HTS.\cite{Alff} The origin of the pseudogap is a subject of
intensive studies. Several authors suggested that this regime is
characterized by the presence of pre-formed Cooper pairs, which
however do not exhibit long range phase coherence.\cite{emery}
Alternate models suggest that the pseudogap region results from
strong superconducting fluctuations,\cite{cucolo}
antiferromagnetic correlations,\cite{chubukov} or separation of
spin and charge.\cite{kotliar} Currently, there is not enough
experimental information to decide which of these models presents
the correct description of the origin of the pseudogap. For
example, one could interpret the microwave spectroscopy
experiments of Corson {\em et al.},\cite{Corson} and the Nernst
effect measurements by Xu {\em et al.},\cite{Xu} as supportive of
the pre-formed pairs scenario. On the other hand, the null results
of experiments searching for Andreev reflections above $T_c$ is
pointing against it.\cite{Dagan,Gonnelli} It is the aim of the
present paper to add experimental information which could be
useful to discriminate between the various models. In our previous
studies we used underdoped YBCO based ramp type (or edge)
junctions, to measure the full angular dependence of the
conductance spectra in the {\it a-b} plane. We found that the
superconducting order parameter shows a modified $d_{x^2-y^2}$
wave behavior with an extended node region and a non zero gap at
the node.\cite{Koren2002} We also investigated the temperature and
magnetic field dependence of the conductance spectra and found
three distinct energy scales, which could be distinguished under
different fields and temperatures. These were attributed to the
two superconducting $d_{x^2-y^2}$ and $is$ (or $id_{xy}$) energy
gaps, and the pseudogap.\cite{KorenJLTP} In the present work we
extend our study to determine the angular dependence of the
pseudogap. We find the same angular dependence for the
superconducting gap and the pseudogap near the surface of
underdoped YBCO as was already found before in BSCO,\cite{Ding}
and discuss some possible reasons of this behavior.\\

Our underdoped YBCO based ramp type junctions with $\rm
YBa_2Fe_{0.45}Cu_{2.55}O_{6+x}$ barrier were described
before.\cite{Koren2002,KorenJLTP,NesherT} Briefly, they consist of
all epitaxial thin film layers of $c-axis$ orientation with
electrodes which are coupled through the junction in the {\it a-b}
plane. The films were prepared by laser ablation deposition, and
patterning was done by deep UV photolithography and Ar ion
milling. The junctions were found to have an extremely smooth
interface of less than a nanometer roughness.\cite{Koren2002} The
multilayer thin film structure was patterned into 10 junctions on
the (100) $\rm SrTiO_3$ wafer, each in a different direction
($\theta$) in the {\em a-b} plane of the films. By the use of
three separate milling steps of the base electrode, similar ramp
angles of $\sim 35\pm 5^\circ$ were obtained for all 10 junctions
on the wafer. Typical film thicknesses are 90 nm for the base and
cover electrodes and 22 nm for the barrier, while the width of
each junction is $5\,\mu m$. A gold layer is deposited on top of
the whole wafer and patterned to produce the 10$\times$4 contact
pads for the 4-terminal transport measurements.\\

Fig. 1 shows the resistance versus temperature of 6 of the 10
junctions on a wafer. Compared to our previous
study,\cite{Koren2002} less oxygen was used in the annealing
process resulting in much more resistive junctions ($\times 5$)
and underdoped YBCO electrodes which become superconducting at
$\sim$50K as compared to 60K used
previously.\cite{Koren2002,NesherT} The normal state resistance is
typical of underdoped $YBa_2Cu_3O_y$ with $6.50\lesssim y\lesssim
6.55$.\cite{Segawa} For junctions with orientation close to the
$a$ or $b$ axes, the resistance goes through a minimum at 40K,
increases to a local maximum at 7K and then decreases at lower
temperature. For junctions with orientation near the node, the
normal resistance $R_N$ stays almost constant between 40 and 20K
and then decreases with temperature down to 7K, where it starts
increasing again slightly.  Similar to our previous
results,\cite{Koren2002} we find that the $R_N$ values of the
$30^\circ$, $45^\circ$ and $60^\circ$ junctions are the lowest.
This result is consistent with the $d_{x^2-y^2}$-wave anisotropy
of the order parameter where states in the gap near the node
region contribute to increase the conductance at low
bias.\\

Fig. 2 shows  the  normalized conductance measured on 3 of the 10
junctions on the wafer at 3.8K, and also at 6.4K for the
45$^\circ$ (node) junction. The absolute values of the conductance
at zero bias can be inferred from the resistance values in Fig. 1
which were measured at low DC bias. Like in our previous
studies,\cite{Koren2002, KorenJLTP} we find a different angular
dependence of the conductance spectra for junctions near the node
orientation (30$^\circ$ and 45$^\circ$) and the one near the main
axis (15$^\circ$). For the sake of simplicity, we take the
measured tunneling gap $\Delta$ of our junctions as the peak to
peak voltage difference divided by 4. The resulting
$\Delta(\theta)$ is shown in the inset of Fig. 2 together with the
gap values measured in our previous study\cite{Koren2002} and the
expected form $\Delta_0|cos(2\theta)|$ where $\rm
\Delta_0=20\,mV$. Both sets of data agree quite well, and the
dominant $d_{x^2-y^2}$-wave symmetry is clearly seen.  We single
out the spectra of the $\theta=45^\circ$ junction,  which exhibits
a zero bias conductance peak at 6.4K, consistent with a
$d_{x^2-y^2}$-wave order parameter. At a lower temperature, this
peak splits, implying the emergence of a secondary order
parameter, previously observed by others.\cite{Laura,Guy} The
finite gap value at $\theta=45^\circ$ in our previous work is well
reproduced\cite{Koren2002, KorenJLTP}, and equals $\rm 3\pm
0.5\,mV$, consistent with $\rm 2.5\pm 0.3\,mV$ value found before.
There is therefore evidence from several sources regarding the
existence of an additional sub-dominant component in the pair
potential of YBCO near a surface (in our case, the interface with
the barrier). Since the gap feature at low bias disappears already
at 6.4K, and a zero bias peak appears as seen in Fig. 2 and also
in Ref. \cite{KorenJLTP}, it cannot result from a finite tunneling
cone which should be independent of temperature.  Therefore, the
observation can be explained as due to a presence of  $is$ or
$id_{xy}$ components in the order parameter near a surface. To
further check this conclusion we also simulated the conductance
curve of the 45$^0$ junction at 3.8K using a $d_{x^2-y^2}+is$ wave
order parameter and a model given by Tanaka, Tanuma and
Kashiwaya.\cite{Irina} The simulation result fits the data very
well, and is shown by the solid line in Fig. 2. Attempts to fit
the data using a $d_{x^2-y^2}+s$ wave order parameter were not
successful. Another feature seen in the normalized conductance of
the node junction in Fig. 2 is that the maximum conductance at low
bias is larger than the Andreev limit of 2 for high transparency
(low Z) junctions. The reason for this is the existence of bound
states also at voltage values higher than zero bias.\cite{Irina}\\

Fig. 3 shows the low temperature conductance spectra of the
$0^\circ$ junction at several magnetic fields normal to the wafer,
together with the corresponding gap values versus field. First we
note that the maxima of the conductance seen in this figure show a
much smaller variation with the field compared to the $45^\circ$
junction in Fig. 2 (17\% versus 250\%). The inset of Fig. 3 shows
the gap apparently increasing with field. We attribute this
apparent behavior to changes in the relative spectral weight of
the superconducting gap and the pseudogap induced by the magnetic
field. With the application of a magnetic field, first the
strongest feature attributed to the $is$ or $id_{xy}$ component of
the order parameter is suppressed, followed by the $d_{x^2-y^2}$
component, finally leaving only the pseudogap feature which
appears at the highest energy. Similar observations were reported
by Krasnov {\em et al} in intrinsic BSCO junctions.\cite{Krasnov}
Their conductance spectra have also shown suppression of
characteristic features with increasing field, while the apparent
gap energies were observed to either decrease or increase with
field depending on temperature. Above $T_c$ in our junctions, the
leads to the junction become normal and their resistance becomes
much larger than that of the junction. Hence, measurements of the
conductance spectra above $T_c$ are not possible. We therefore
used the magnetic field to suppress superconductivity at low
temperature, as was done recently also by Alff {\em et
al.}.\cite{Alff} In our previous observations, the apparent peak
position of the conductance in the 0$^\circ$ junction was found to
decrease with increasing field up to 5T.\cite{Koren2002} The
reason for the difference from the present results can be due to
the fact that the present junctions are much more resistive and
thus more in the tunneling-like regime. The inset of Fig. 3 shows
that upon increasing the field up to 1.1T, the position of the
peak in the conductance spectra increases from 9 to 20 mV with a
plateau at $\sim$19 mV which is similar to the d-gap value of the
60K YBCO phase measured previously ($16\pm 1.5$
mV).\cite{NesherAndreev} This effect is attributed to the
suppression of the $is$ component whose signature disappears
already at $\sim$0.2T. At higher fields, the $d_{x^2-y^2}$ gap
signature in the conductance is also suppressed (at $\sim$1.4T),
and the conductance spectra exhibit only a peak around 110 mV
which corresponds to a gap value of 55 mV. This value is
characteristic of the pseudogap in underdoped YBCO, and we
therefore identify this feature as the pseudogap.\cite{Racah}
At 8T, the peak position again decreases slightly.\\

Why should a field in the range of a few Tesla (small compared to
$H_{c2}$) suppress the $d_{x^2-y^2}$ gap signature (of the 50K
YBCO phase) at 4K is puzzling, but the large 50-60 mV gap values
observed here are clearly due to the pseudogap and not the
$d_{x^2-y^2}$ gap.\cite{Racah} We suggest the possibility that
superconductivity near the interface is weakened by the proximity
effect with the normal barrier. This effect is apparently much
stronger in our new, more resistive junctions. Preliminary results
using junctions with a {\em non-magnetic} Ga doped YBCO barrier,
also show similar behavior in magnetic fields as observed in the
present study in junctions with a Fe doped YBCO barrier. It is
therefore quite unlikely that the depression results from
scattering in the barrier, as the result seems independent of the
type of dopant used in the tunneling barrier. We also note that
conductance spectra in which the relative spectral weight of the
pseudogap feature is larger than the d-gap were observed before,
even at zero field.\cite{DeutscherNature,Racah} Thus the
$d_{x^2-y^2}$-wave characteristic is not fully suppresses at 1-2T,
but only becomes weaker in comparison with the pseudogap.
Recently, an $H_{c2}$ value of about 35T was measured for
underdoped YBCO with $T_c$=50K like we use here.\cite{Segawa} This
relatively low $H_{c2}$ value together with a weaker
superconductivity near the interface, can also help explain the
surprisingly strong suppression of the d-wave characteristic.
Going back now to the results in the inset of Fig. 3, one can
deduce that the ratio of critical fields
$H_{c2}(d_{x^2-y^2})/H_{c2}(is)$ in underdoped YBCO is of the
order of 1.4/0.2=7.  Taking $H_{c2}(d_{x^2-y^2})$ as 35T, we can
estimate that the critical field of the $is$ component of the
order parameter should be of about 5T.\\

Fig. 4 shows the conductance spectra of the $45^\circ$ junction at
4K under several magnetic fields normal to the wafer, together
with the corresponding gap values. As discussed before, since a
pure $d_{x^2-y^2}$-wave order parameter has no gap along the node,
the prominent features at small fields in the conductance
characteristics of Fig. 4 must be due to the small $is$-wave
component.\cite{Koren2002,KorenJLTP} We find that with increasing
field, the whole conductance spectrum is almost totally
suppressed, similarly to what was observed with the $0^\circ$
junction. We attribute the non conservation of spectral weight in
the normalized conductance spectra versus field to the breaking of
bound states which are prevalent at energies $E \lesssim
2\Delta_d$.\cite{Irina} The position of the conductance peak
increases significantly, from 3 mV at 0T to 7.5 mV at 8T as seen
in the inset of this figure. As was explained above, the larger
gap values seen at high fields (see inset of Fig. 3) are due to
the pseudogap. Therefore, the conductance characteristics at high
fields in Fig. 4 must also be due to the pseudogap. It seems that
by applying an 8T field, we can apparently "erase" all other
features except the pseudogap. We are therefore in the position to
perform a clean measurement of the angular dependence of the
pseudogap in underdoped YBCO.\\

Fig. 5 presents the main message of this study. It shows the
normalized conductance spectra at 4K and 8T of six junctions with
different orientations in the {\it a-b} plane (main panel), and
the resulting angular dependence of the measured pseudogap feature
(inset). We find that the angular dependence of the pseudogap is
the same as that of the superconducting gap given in the inset of
Fig. 2 (to within experimental error). This indicates that the two
phenomena might have a common origin. The most obvious reason is
that both are related to the intrinsic crystalline symmetry of
YBCO. It would appear that since the signature of the
superconducting gap disappears with the application of a magnetic
field and that of the pseudogap does not, the pseudogap is not
connected with superconductivity. However, since the value of the
magnetic field needed to suppress these features seems to depend
on the energy of the peak in the conductance curve, it could be
that the field we have at our disposal (8T) is simply insufficient
for the suppression of the pseudogap feature. Hence, other
scenarios which are consistent with the present results cannot be
ruled out . One such scenario is that of the possible existence of
uncorrelated pre-formed pairs in the pseudogap
regime.\cite{Buchanan} These pairs have the same symmetry of the
pair-wave function as that of the correlated pairs and can thus
naturally yield the same symmetry for the pseudogap
and the superconducting gap.\\

It should be noted though that angular resolved photoemission
measurements (ARPES) of the angular dependence of the gap and the
pseudogap have been done before in BSCO 2212 by Ding {\em et
al.}.\cite{Ding} They found that there is basically no significant
difference between the observed gap below and above $T_c$. The
magnitude and angular dependence of the gap and pseudogap was
almost the same, except for the observation of a broader node
region for the pseudogap. The error in the ARPES measurements
however, in the values of the energy gap of the heavily underdoped
BSCO was quite large, about $\pm 60\%$. In contrast, the error in
the present conductance measurements in the determination of the
energy gap values near the node region is much smaller, about $\pm
20\%$. Therefore, the added value of the present results is the
higher energy resolution, the use of a different (tunneling)
technique, and the fact that we study another material, underdoped YBCO.\\

In conclusion, our results of the angular dependence of the
conductance demonstrate clearly that both the superconducting gap
and the pseudogap have the same $|d_{x^2-y^2}+is|$-like symmetry
in underdoped YBCO near the interface of the junctions. This
result, combined with the similar result found for underdoped
BSCCO,\cite{Ding} imposes another constraint on the theoretical
description of the pseudogap. We also found that $H_{c2}$ of the
small $is$ component of the order parameter is of the order of 5T.\\

We are grateful to A. Auerbach, O. Millo, G. Deutscher and I.
Lubimova for useful discussions. We also thank I. Lubimova for use
of her conductance simulation program. This research was supported
in part by the Israel Science Foundation, the Heinrich Hertz
Minerva Center for HTS, the Karl Stoll Chair in advanced
materials, and by the Fund for the Promotion of Research at the Technion.\\

\newpage

{\bf \large Figure Captions}
\begin{description}

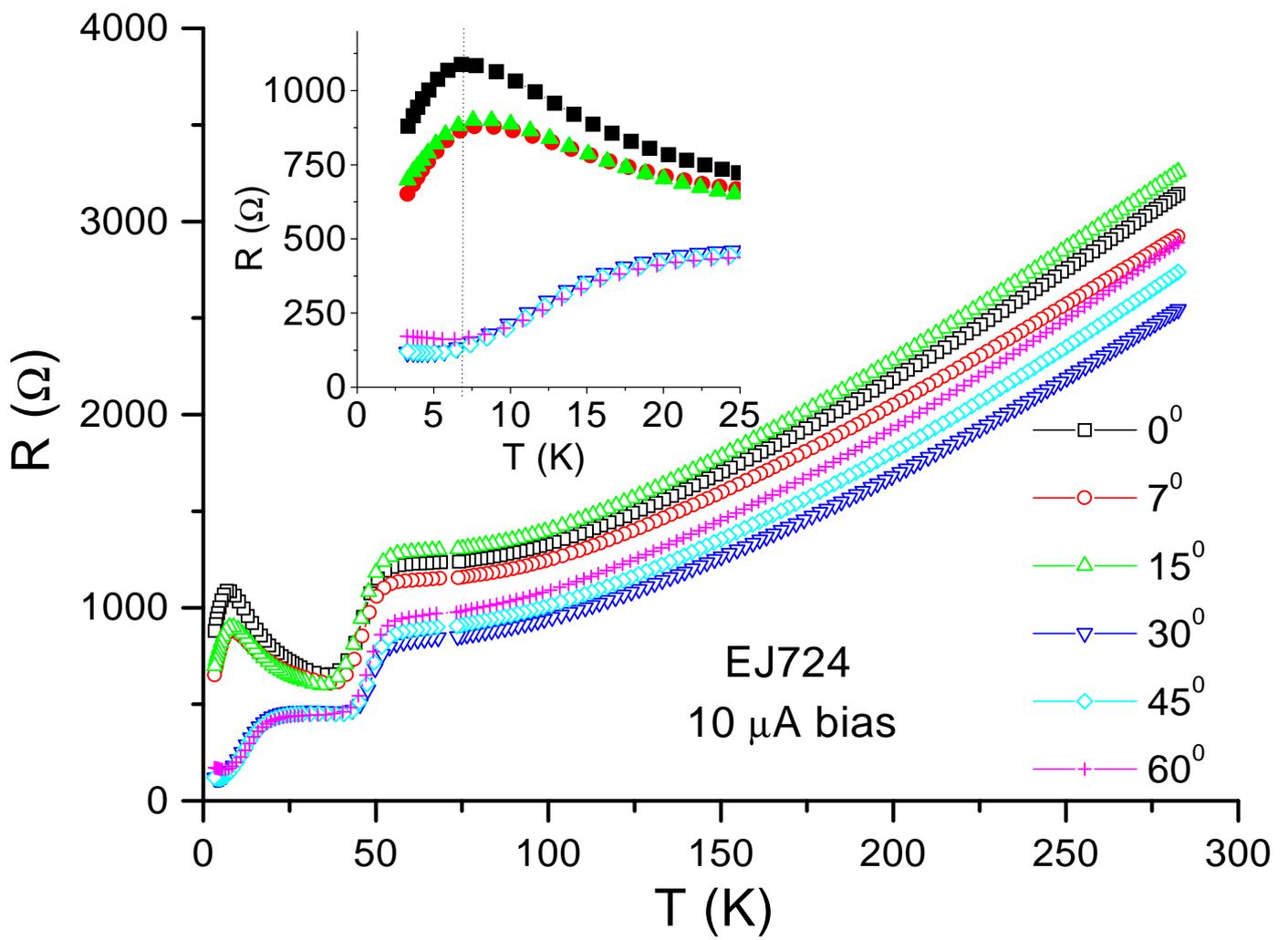
\item [Fig. 1:] Resistance versus temperature of six
of the ten junctions on the wafer. The inset shows the low
temperature region, illuminating the difference between junctions
along the main axes (solid symbols) and near the node (open
symbols).

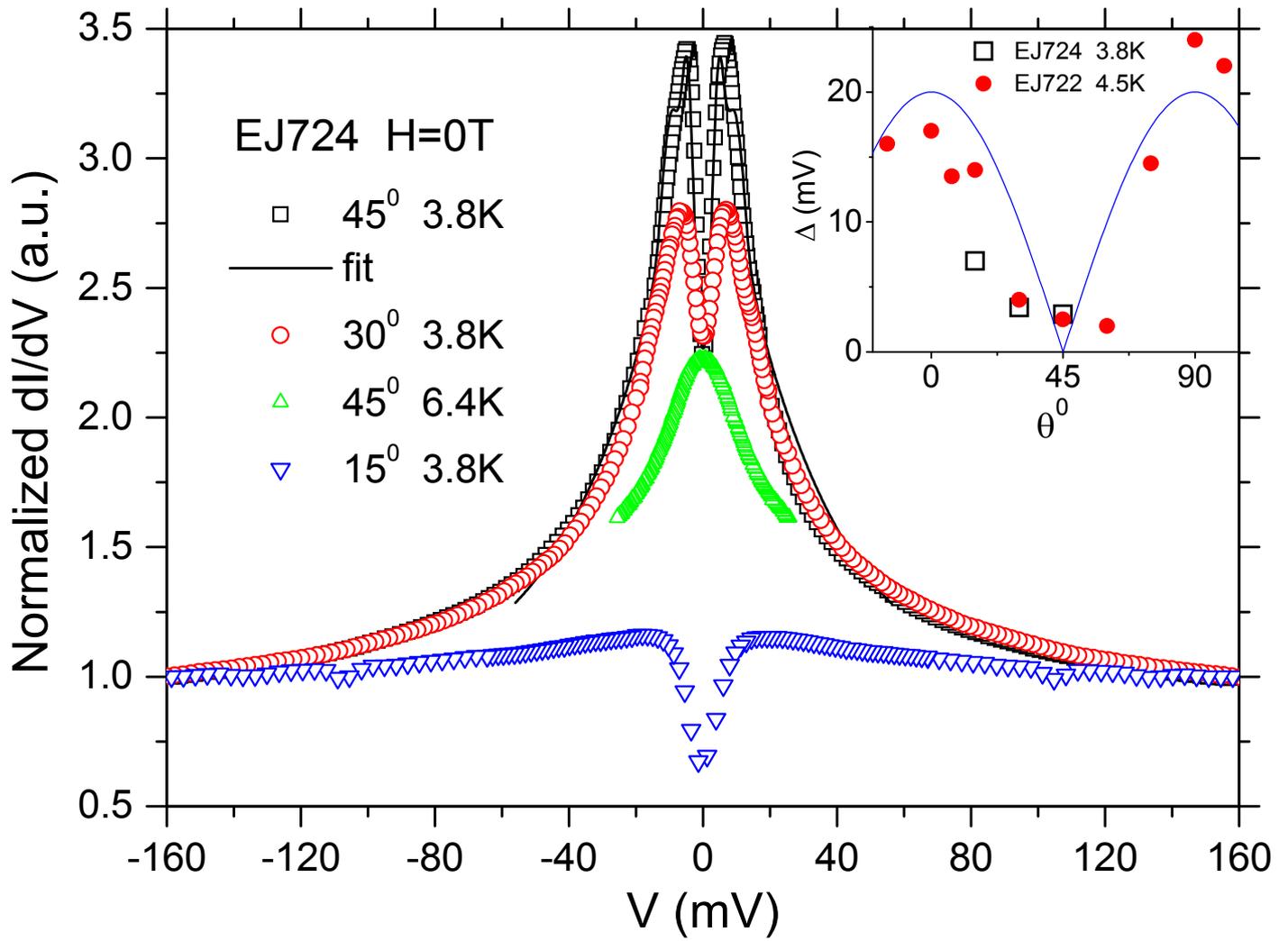
\item [Fig. 2:] Normalized conductance versus voltage
bias of three of the junctions, together with a simulation of the
node junction data at 3.8K (line). The inset shows the angular
dependence of the gap of these junctions at 3.8K (open squares),
together with previous gap values taken from Ref. \cite{Koren2002}
at 4.5K (solid circles), and 20$|cos(2\theta)|$ (line).

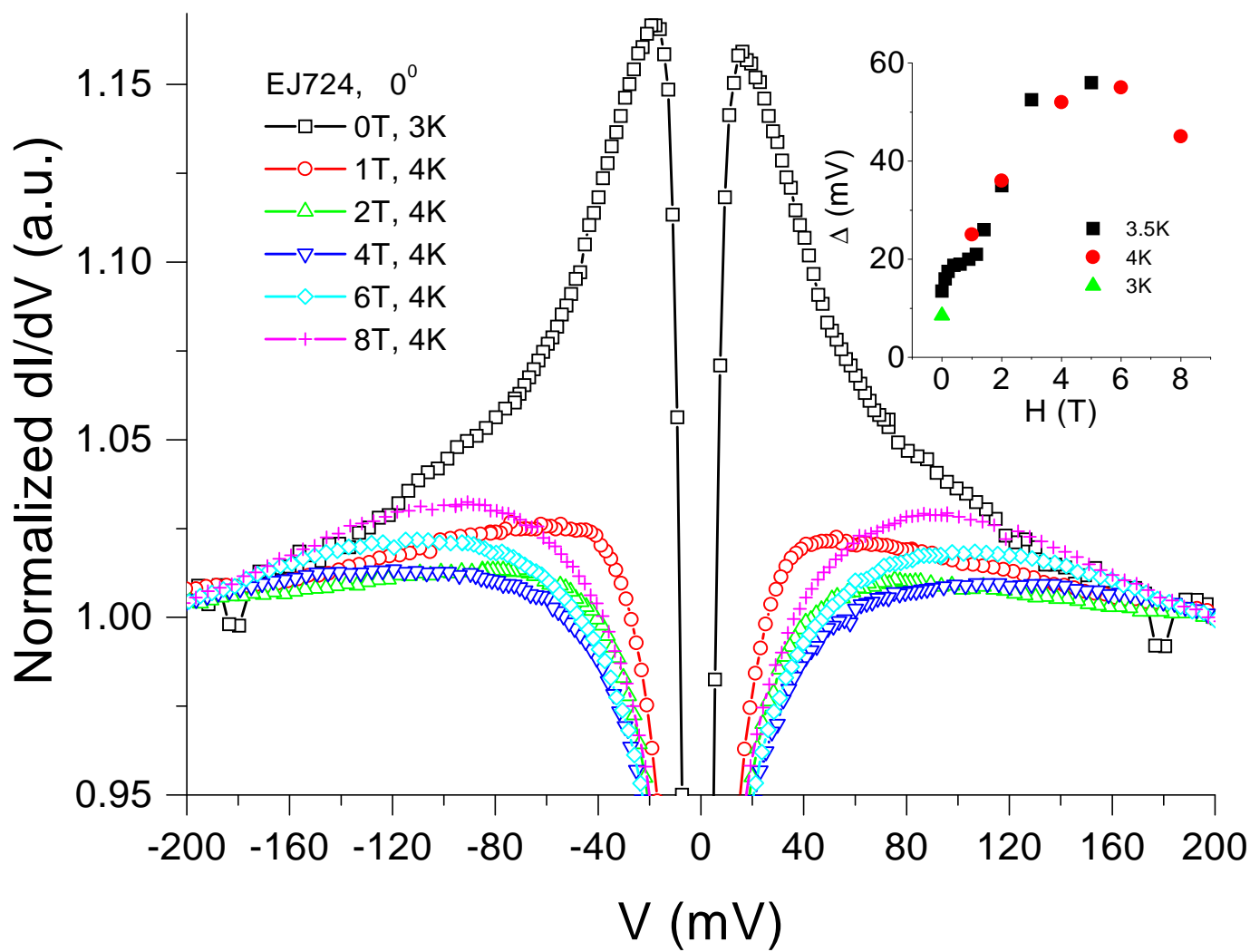
\item [Fig. 3:] Conductance spectra  at low
temperatures of the 0$^\circ$ junction  for several magnetic field
values (main panel), and the corresponding gap energies versus
field (circles and a triangle) together with more data (squares)
from Ref. \cite{KorenJLTP} (inset).

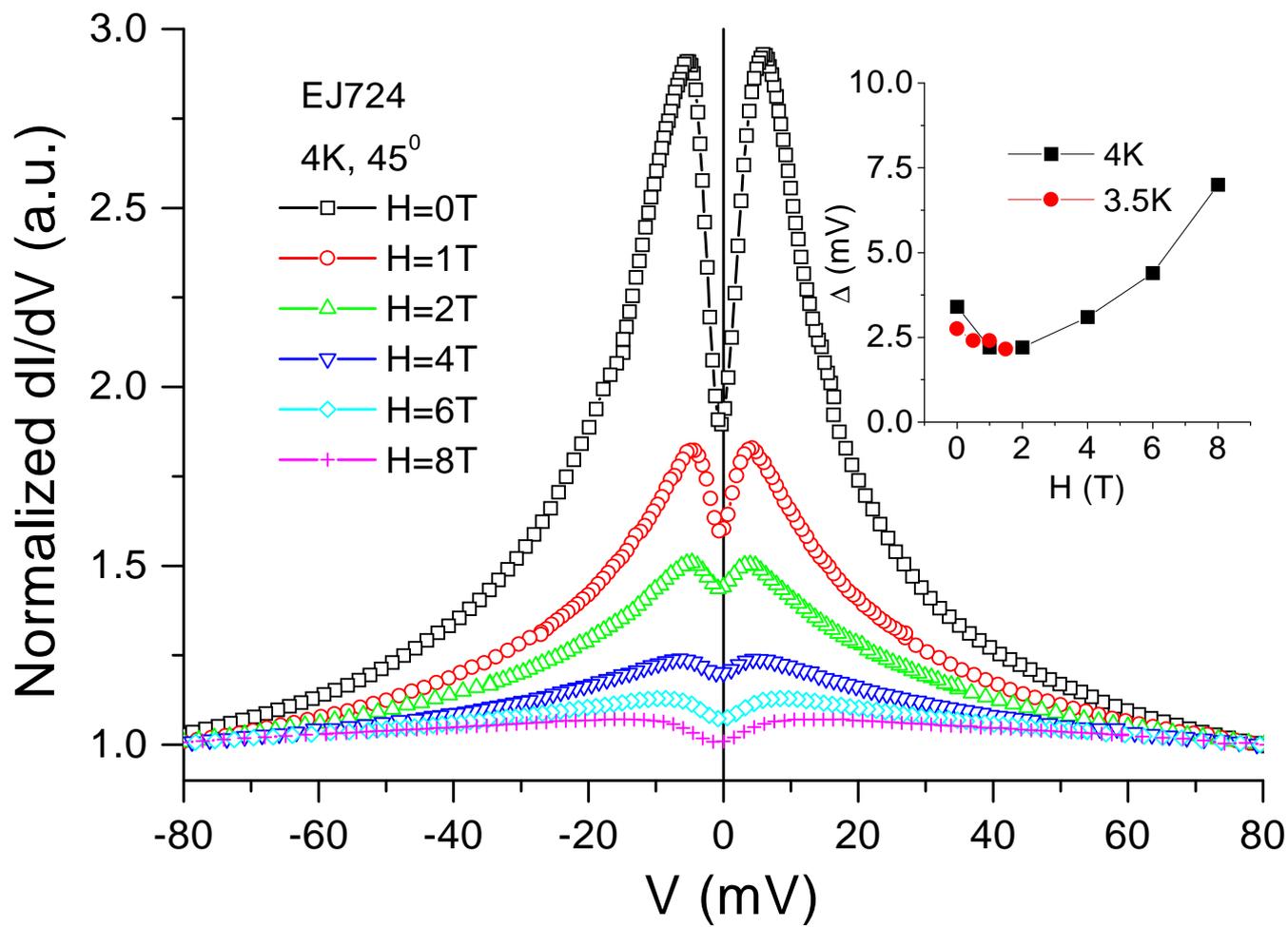
\item [Fig. 4:] Conductance spectra  at 4K of the
45$^\circ$ junction at several magnetic fields (main panel), and
the field dependence of the corresponding gap energies together
with more data at 3.5K (inset).

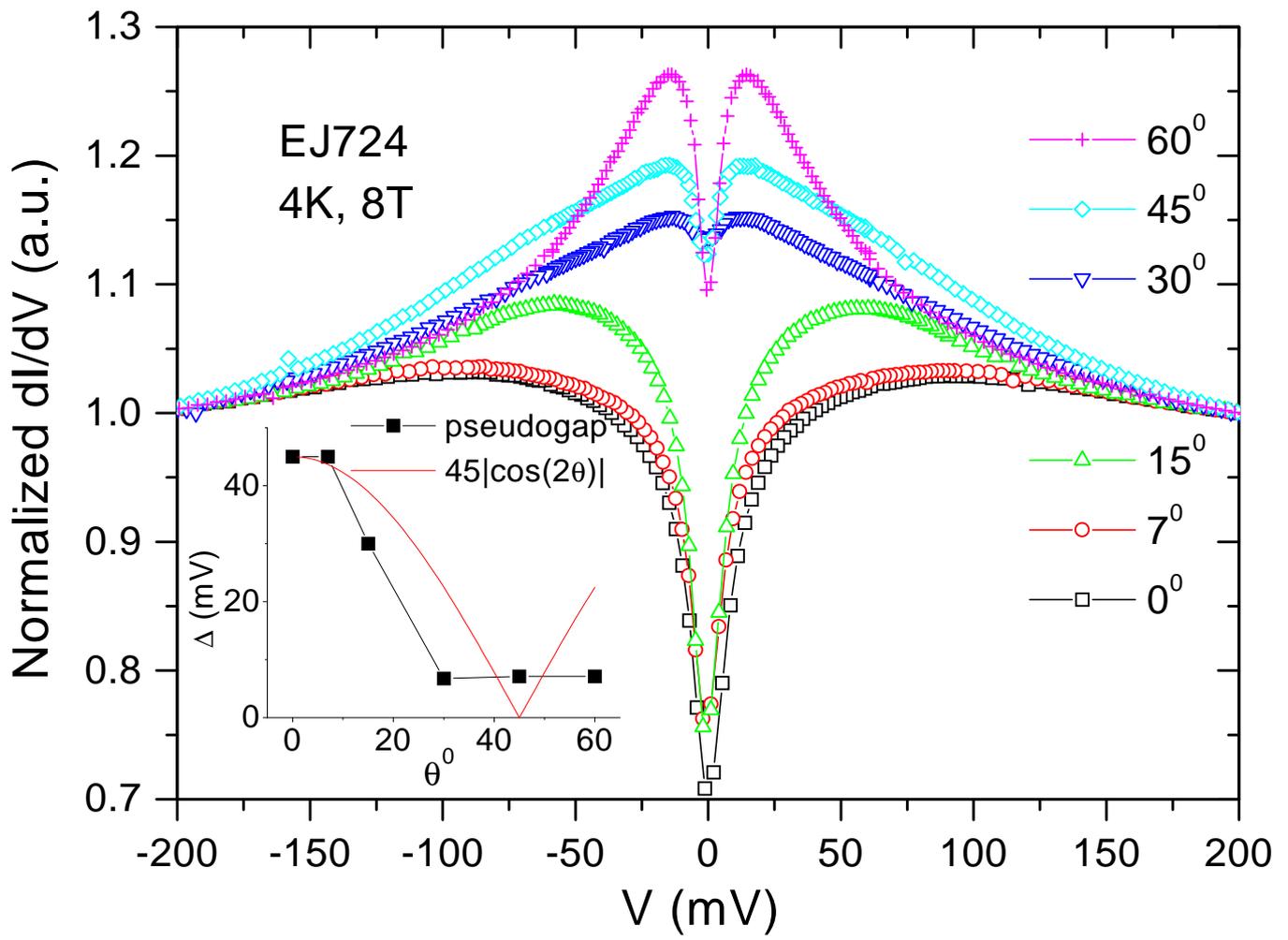
\item [Fig. 5:] Normalized conductance spectra  at 4K
and 8T of six junctions (main panel), and the angular dependence
of the pseudogap and 45$|cos(2\theta)|$ (inset).

\end{description}
\newpage

\bibliography{AndDepBib.bib}

\end{document}